\documentclass[12pt]{iopart}
\usepackage{graphicx}
\begin{document}

\title[Importance of bulk states for the electronic structure of semiconductor surfaces]{Importance of bulk states for the electronic structure of semiconductor surfaces: implications for finite slabs}

\author{Keisuke Sagisaka$^{1,2}$, Jun Nara$^3$, David Bowler$^4$}

\address{$^1$Research Center for Advanced Measurement and Characterization, National Institute for Materials Science\\1-2-1 Sengen, Tsukuba, Ibaraki, 305-0047 Japan}
\address{$^2$London Centre for Nanotechnology, 17-19 Gordon St, London, WC1H 0AH, U.K.}
\address{$^3$International Centre for Materials Nanoarchitectonics (MANA), National Institute for Materials Science\\1-1 Namiki, Tsukuba, Ibaraki, 305-0044 Japan}
\address{$^4$Thomas Young Centre and Department of Physics $\&$ Astronomy, University College London\\Gower St, London, WC1E 6BT, U.K}

\ead{SAGISAKA.Keisuke@nims.go.jp, david.bowler@ucl.ac.uk}
\vspace{10pt}

\begin{abstract}
We investigate the influence of slab thickness on the electronic structure of the Si(100)-\textit{p}(2$\times$2) surface in density functional theory (DFT) calculations, considering both density of states and band structure. Our calculations, with slab thicknesses of up to 78 atomic layers, reveal that the slab thickness profoundly affects the surface band structure, particularly the dangling bond states of the silicon dimers near the Fermi level. We find that, to precisely reproduce the surface bands, the slab thickness needs to be large enough to completely converge the bulk bands in the slab. In case of the Si(100) surface, the dispersion features of the surface bands, such as the band shape and width, converge when the slab thickness is larger than 30 layers. Complete convergence of both the surface and bulk bands in the slab is only achieved when the slab thickness is greater than 60 layers. 
\end{abstract}

%
\noindent{\it Keywords}: density functional theory, band structure, semiconductor surface, silicon, slab thickness
%
%
\maketitle
%
%

\section{Introduction}

Semiconductor surfaces are of enormous importance both industrially, as the basis of the microelectronics industry, and scientifically, where they form the basis for studies of diffusion, growth and electronic structure.  Moreover, recently a number of proposals for quantum information implementations based on isolated states in or near semiconductor surfaces have emerged \cite{Veldhorst15}.  In light of this, it is vital that our understanding of their atomic and electronic structure is as accurate and detailed as possible.  We demonstrate in this paper that both the gap and the character of the valence band depend strongly on the thickness of slabs used to model the system, and find that the slabs used in most calculations are much too small to describe the electronic structure of Si(100), the most common surface in industry.

Density functional theory (DFT) has become the standard approach to model the atomic and electronic structures of semiconductor surfaces. Although it is known to underestimate the band gap of solids, DFT generally reproduces the density of states (DOS) and band structure accurately. A surface is usually modeled by using a supercell consisting of a slab with two surfaces and a vacuum region. Since this approach gives a non-periodic system perpendicular to the surface (z direction), the accuracy of the resulting electronic structure will depend on the slab thickness. In case of the Ge(100) surface, which bears a very similar structure to Si(100), the influence of slab thickness on the top of the valence band has been discussed in detail \cite{Radny08, Yan09, Radny09, Shah12}. Some other physical properties, such as surface electronic structure \cite{Smeu12}, dielectric function \cite{Vazhappilly14}, chemical reactivity \cite{Shah12}, and transport properties \cite{Martinez09}, sensitively vary with cell size and slab thickness. Since recent improvements to experimental techniques allow measurements to be made with high energy resolution ($\sim$ meV), slab thickness is likely to become a key parameter in determining energy precision in DFT modeling.

However, it is still unclear how thick a slab is needed to reliably produce the electronic structure of a semiconductor surface; one study \cite{Shah12} suggested that slab thickness was unimportant, while another study of the basic $\left( 2\times 1\right)$ reconstruction \cite{Seo14} suggested, without giving detailed analysis, that at least 30 layers are required. Recently, we found that previously reported band structures of Si(100) calculated with DFT \cite{Fritsch95, Zhu89, Ramstad95} do not compare well with data obtained by the state-of-the-art angle-resolved photo-emission spectroscope (ARPES) \cite{Takayama15}. In addressing this problem, we found that surprisingly thick slabs were required to obtain agreement with the ARPES data. In this paper, we report the critical slab thicknesses in the DFT calculation required to reliably produce the electronic structure of the Si(100) surface in terms of DOS and band structure. 

\section{Computational method}
DFT calculations were carried out using the VASP code with a plane wave basis set \cite{Kresse93, Kresse96}. We employed PAW and the GGA PBE exchange correlation functional \cite{Perdew96}. The Si(100)-\textit{p}(2$\times$2) surface was modeled using supercells consisting of a Si slab and vacuum layer of approximately 15 \AA\ thick (this vacuum thickness was carefully converged). The bottom layer was terminated with hydrogen atoms and the bottom layer and hydrogen layer were fixed; as the system is not polar, and the bottom surface is inert, no dipole corrections are required.  For total energies and DOS, the tetrahedron method was applied \cite{Blochl94}. Full details are given below.

\paragraph{Parameters used}
We tested the influence of slab thickness between 6 and 78 atomic layers (L) on the electronic structure of the slab, in particular the dangling bond (DB) states.  To ensure full convergence with respect to the basis set, we employed a plane wave cutoff energy of 312.5 eV throughout this study. Structural relaxation was done with Brillouin zone (BZ) sampling of 8$\times$8$\times$1 Monkhorst-Pack \textbf{k}-point mesh until the forces on each atom reached to below 0.02 eV/\AA. The relaxed lattice constant was $a_0$ = 5.466 \AA, which is slightly larger than experimental value (5.43 \AA). For the total energies and DOS plots, the BZ sampling was increased to 22$\times$22$\times$1. 

\paragraph{Surface energies}
Surface energies were calculated using both a reconstructed slab and a
fully hydrogen terminated slab.  The total energies were calculated
for the p(2$\times$2) slabs (the top was dimerised, the bottom
was terminated with hydrogen) and hydrogen terminated slabs (both
surfaces were terminated with hydrogen, using the same structure as the
slab base) for thicknesses of 6 L, 14 L, 22 L, 30 L, 38 L, 62 L and 78 L.
All simulation cell sizes were constant (2$\times$2$\times$78L). As the slab
thickness was varied, so the amount of vacuum space also varied.
Each p(2$\times$2) slab was relaxed with a 8$\times$8$\times$1 k-point sampling, then the
total energy was calculated with a 22$\times$22$\times$1 k-point
sampling, with the  tetrahedron method.  By contrast, each hydrogen
terminated slab was unrelaxed. The total energy was calculated with 
a 22$\times$22$\times$1 k-point sampling, again using the tetrahedron
method.  For all these calculations, the plane wave cutoff was increased to 600 eV.

To find a bulk silicon atomic energy, the total energy was calculated
for a periodic 2$\times$2$\times$4 cell (16 Si atoms, bulk
calculation), which was then divided by 16 to give the energy of a
Si atom.

Then, the surface energy was found by subtract the energy for hydrogen
termination (for one surface) and the energy of the bulk silicon atoms
from the energy for the p(2$\times$2) cell, divided by 2 to give eV/dimer.

\paragraph{Band Structure}
For band structures and densities of states, we performed a static
calculation with 22$\times$22$\times$1 k-point sampling to
produce a charge density. (4$\times$4$\times$1 for the hybrid calculation)

To find a consistent energy zero for comparison between different slab
thicknesses, all eigen values were shifted relative to the potential at the Si core, averaged
over the first five layers below the surface. The top of the valence
band of the 2$\times$1 symmetric 38L in Fig. 4(d) was arbitrarily chosen
as the zero for all figures. All bands in Fig. 4 in the main text are
drawn with respect to a consistent energy zero, the potential value at the Si cores.

\paragraph{Bulk band gap for a slab}
The bulk band gap of a slab was found by considering the projected DOS of Si atoms in
the middle layer of the slab.  We found that it converged for 62 L
(interestingly, when considering a periodic bulk two atom cell, we
required a k-point mesh 31$\times$31$\times$31, giving equivalent sampling).

\paragraph{Hybrid functional band structure}
For the hybrid functional calculations, owing to the large
computational cost, the charge density was found self-consistently
using a 4$\times$4$\times$1 k-point mesh.

To start, we tested the influence of slab thickness on three key properties: the Si dimer bond length and buckling angle; and the surface energy. The structural properties are relatively weakly dependent on the slab thickness (Table \ref{tab:table1}): the bond length and buckling angle both converge at 14 L.  Performing calculations with a slab of less than 14 L is probably acceptable for the study of surface geometry, but will require some care. On the other hand, surface energy convergence is only obtained when the slab thickness is greater than 22 L, with the surface energy for a 14 L thick slab 114 meV \emph{per dimer} higher than the converged value. We recall the discussion of the relative stability of the \textit{c}(4$\times$2) and \textit{p}(2$\times$2) reconstructions \cite{Fritsch95,Inoue94,Nakamura05}, which involved energy differences of a few meV, for which the error in surface energy, even at 14 L, is far from negligible. For calculations considering reaction, adsorption, or diffusion at the Si(100) surface, slab thickness needs to be tested and chosen carefully. 

\begin{table}
\caption{\label{tab:table1} Dimer bond length, dimer buckling angle, and surface energy obtained for the Si(100) \textit{p}(2$\times$2) surface with varying slab thicknesses. For surface energy calculations, we used a constant cell size (7.731$\times$7.731$\times$121.914 \AA $^3$) with different slab thicknesses and increased the cutoff energy up to 600 eV.} 
\footnotesize
\begin{tabular}{@{}llll}
\br
	Slab thickness (layers)& Dimer bond length (\AA) & Buckling angle ($^\circ$)& Surface energy(eV/dimer)\\
\mr
	6   & 2.360 & 19.0 &  2.424\\
	14 & 2.357 & 19.1 &  2.446 \\
	22 & 2.355 & 18.9 &  2.331 \\
	30 & 2.355 & 19.1 &  2.331 \\
	38 & 2.357 & 19.1 &  2.332 \\
	62 & 2.357 & 19.1 &  2.333 \\
	78 & 2.357 & 19.0 &  2.332 \\
\br
\end{tabular}

\end{table}
\normalsize

\begin{figure}
\includegraphics[width=13cm]{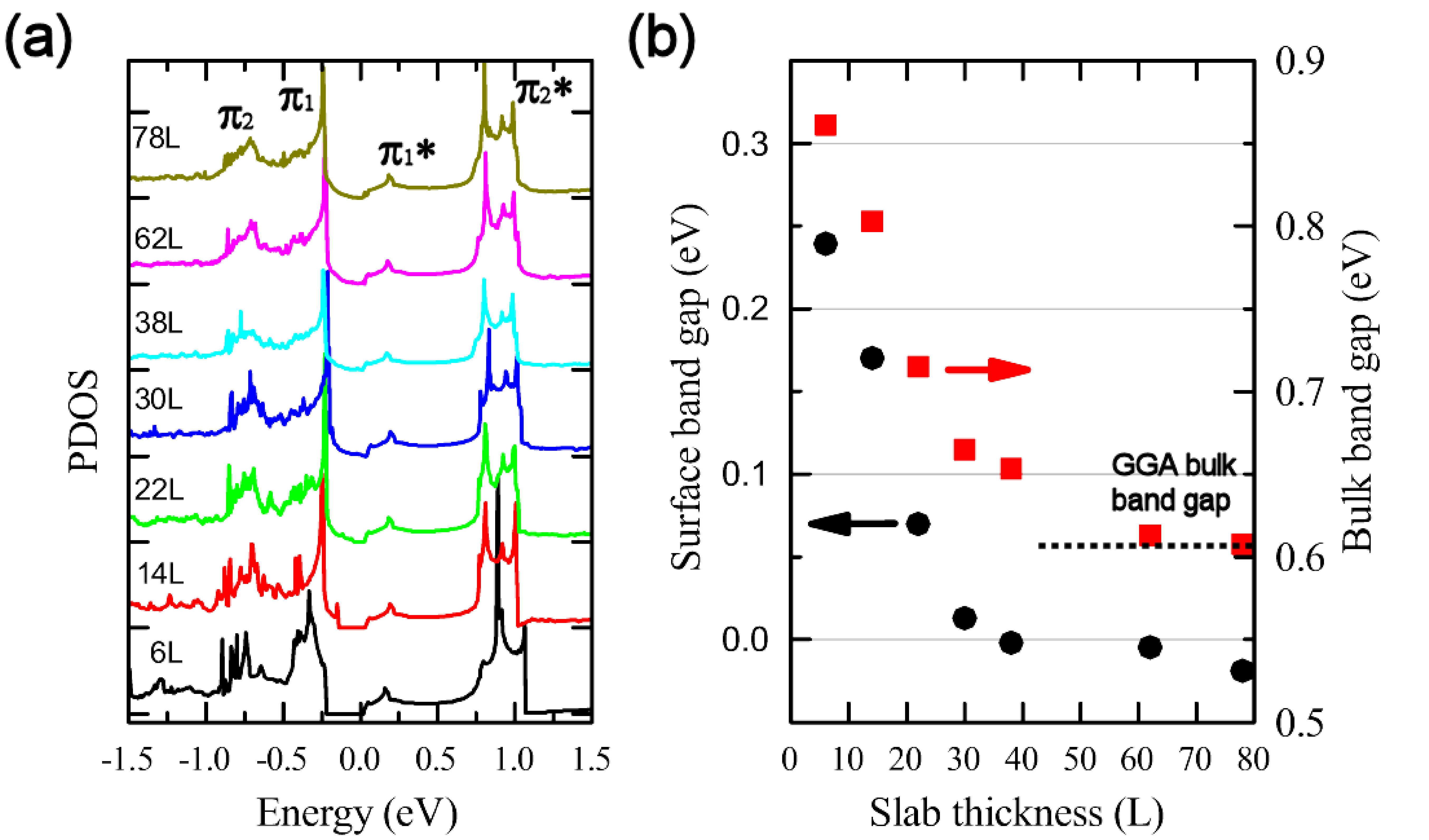} 
\centering
\caption{(Color online) (a) PDOS of Si dimer calculated with Si(100) \textit{p}(2$\times$2) slabs. All energies are shifted so that the bottom of the potential of Si atoms averaged over the top five layer of each slab has the same value. (b) Surface (circle) and bulk (square) band gaps as a function of the slab thickness. The bulk band gap sizes were found from PDOS of the middle layer in each slab. The GGA bulk band gap value obtained by a converged bulk calculation is indicated by a broken line.}
\label{fig:PDOS}
\end{figure}

We now turn to the key electronic structure: DOS and the band gap, on which the slab thickness has a profound effect. Figure \ref{fig:PDOS}(a) shows the DOS projected onto a Si dimer for slab thicknesses from 6 L to 78 L. The common features among these DOS are two peaks in the filled states, and one broad and one pronounced peak in the empty states: these are often referred to as $\pi_1$,$\pi_2$, $\pi_1^*$ and $\pi_2^*$, as labeled in Fig. \ref{fig:PDOS}(a) \cite{Ramstad95, Okada01}.  Their appearance agrees well with scanning tunneling spectroscopy results (See Fig.~\ref{fig:STSvsDFT}). The key effect of increasing slab thickness that can be seen is a change in the shape and character of the valence band maximum (VBM). As will become clear below, the VBM of thin slabs are dominated by the DB states, while those of thicker slabs are dominated by bulk states. 
To the best of our knowledge, the VBM character of a clean Si(100) surface has not been well studied yet, and we will discuss this based on the results of band calculations below. 

The second major effect of the increasing slab thickness is a reduction of the surface band gap. Figure. \ref{fig:PDOS}(b) shows the surface and bulk band gaps (gap at the center of the slab) as a function of the slab thickness.
The reduction of the surface band gap with slab thickness is strongly correlated with the reduction of the bulk band gap, and the development of the bulk bands. When the slab thickness is 38 L or greater, the bottom of the $\pi_1^*$ state overlaps with the top of the valence band, which has become a bulk state, and the surface band gap closes.  
It is maybe not surprising that there is some dependence of gap on slab thickness, as, a slab will behave as a form of quantum well, with a thin slab strongly confining its eigenstates, giving a larger band gap \cite{Delley95}. 
In our work, we calculated the fully converged bulk band gap of silicon to be $\sim$0.607 eV, which is only reached when a slab calculation with 62 L or more is performed. 

\begin{figure}
\includegraphics[width=16cm]{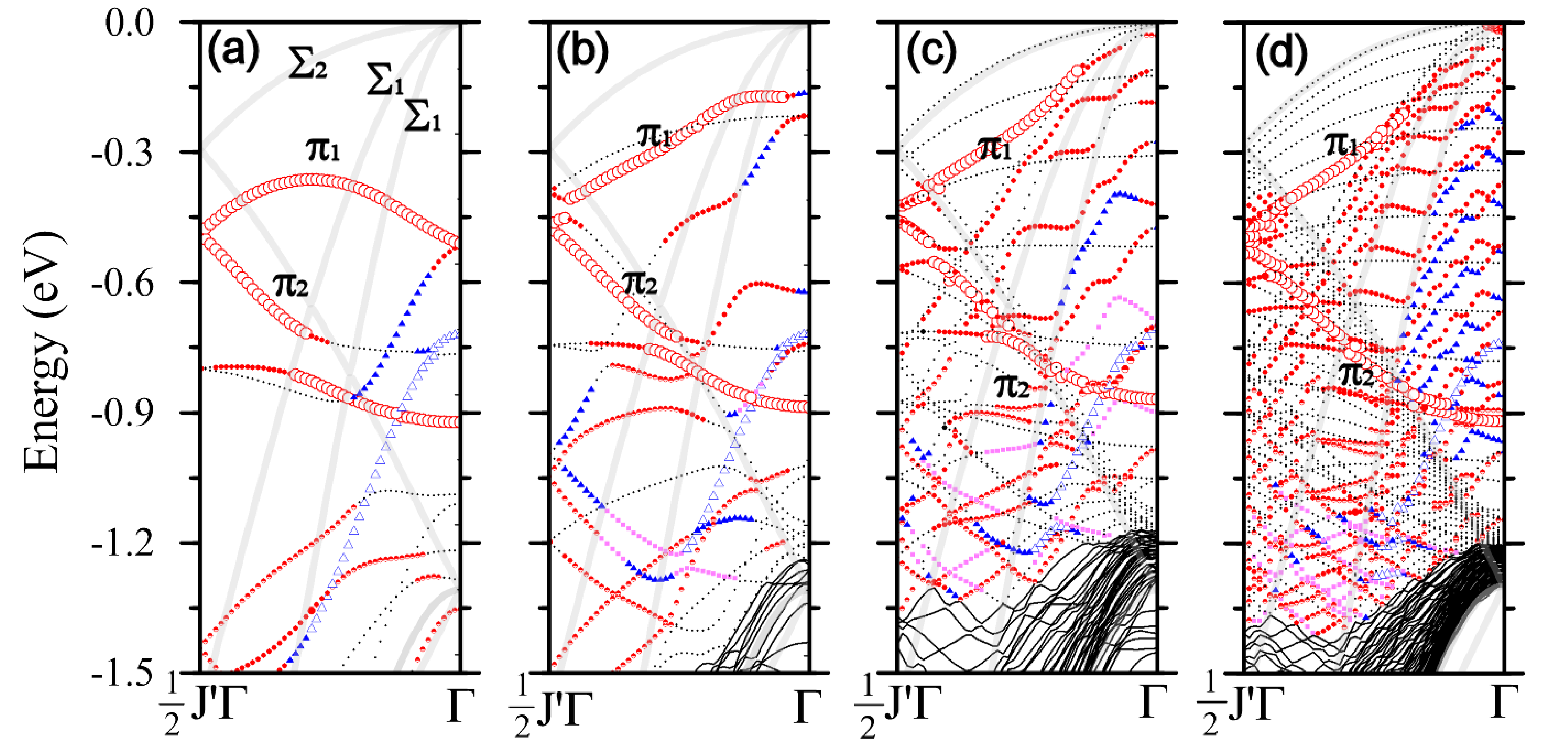} 
\centering
\caption{(Color online) Band structures for the Si(100) \textit{p}(2$\times$2) surface along the dimer row direction ($\Gamma$-$\frac{1}{2}$J$^{\prime}\Gamma$ line). Slab thickness used are (a) 6 L, (b)14 L, (c) 38 L, and (d) 78 L. All bands are drawn with reference to the bottom of the potential of Si atoms averaged over the top five layer of each slab. Red open circles represent eigenstates with intense and localized charges on DB assigned by their charge values being larger than (a) 10, (b) 5.0, (c) 2.0, and (d) 1.0 $\times$ 10 $^{-3}$ e/\AA $^3$; Red dots: hybrid states of DB and bulk states whose charge density is less than the foregoing criteria; Red half filled circles: hybrid states of DB, db and bulk states; Blue open triangles: intense and localized BB state; Blue solid triangles: hybrid states of BB and bulk states; Pink squares: hybrid states of BB, db and bulk states; Black dots: pure bulk states. Solid lines are unchecked bands. Gray lines are bulk bands (at $k_z$ = 0) obtained separately by a bulk calculation with a \textit{p}(1$\times$1) 4 atom cell.}
\label{fig:Band}
\end{figure}

\begin{figure}
\includegraphics[width=13cm]{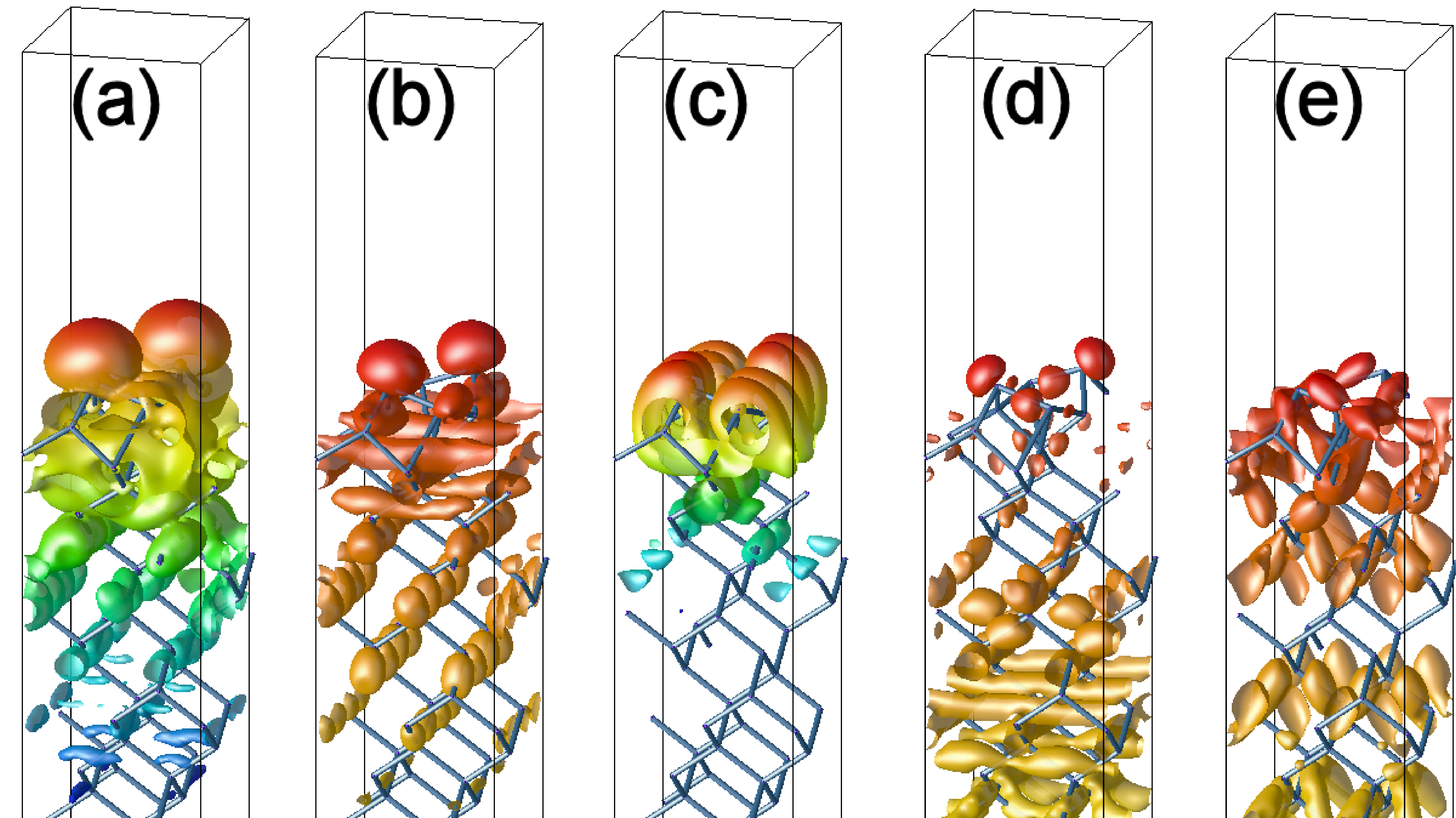} 
\centering
\caption{(Color online) Typical isosurface maps of charge density of the individual eigenstate calculated with a Si(100) \textit{p}(2 $\times$ 2)-78 L slab. (a) DB state ($\bigcirc$), (b) DB + bulk hybrid state ($\bullet$), (c) BB state ($\bigtriangleup$), (d) DB + db + bulk hybrid state ($\ominus$), (e) BB + db + bulk hybrid state ($\opensquare$). Isosurface value: 1.0 $\times$ 10$^{-3}$ e/\AA $^{-3}$. Color scale represents height in z direction. Each symbols are used in Fig. \ref{fig:Band} to group types of states.}
\label{fig:PARCHG}
\end{figure}

In order to gain further insights into the influence of the slab thickness on the electronic structure, we have examined the band structure of the Si(100)-\textit{p}(2$\times$2) surface with various slab thicknesses. Figure \ref{fig:Band} shows valence band 
diagrams near the Fermi level along the dimer row direction ($\Gamma$-$\frac{1}{2}$J'$\Gamma$ line). 
We plotted the partial charge densities for each eigenstate at different k-points in the Brillouin zone (BZ) to assign the character (bulk, DB, back-bond) to the band as it moves through the BZ. Typical isosurfaces of the charge density for the 78 L slab are shown in Fig. \ref{fig:PARCHG} and in Fig.~\ref{fig:PARCHG78}. The major surface states are the DB states $\pi_1$ and $\pi_2$ (red open circles). Both the $\pi_1$ and $\pi_2$ bands show intense charge densities on the Si dimer atoms and they decay within a few layers of the surface. The $\pi_1$ state bears charges localized on the up atom of the Si dimer, while the $\pi_2$ state has most of charges on the up atom and some extended onto the down atom as well.

We observe that the band shape and width of $\pi_1$ drastically changes with the increase of the slab thickness, while those of $\pi_2$ are only slightly changed by the slab thickness. The surface band maximum (SBM) appears half way along the $\Gamma$-$\frac{1}{2}$J$^{\prime}\Gamma$ line for 6 L. Correspondingly, the $\pi_1$ band width is only 0.13 eV. With increasing thickness, the SBM moves toward the $\Gamma$ point and the band width also widens. Eventually, the SBM reaches to the $\Gamma$ point when the slab thickness is greater than 20 L and the band width converges, to approximately 0.40 eV, for thickness greater than 30 L (see Table~\ref{tab:table2}).

The convergence of the surface states relies on more than just slab thickness: it also requires precise reproduction of the bulk electronic structure in the slab. In Fig. \ref{fig:Band}, the eigenstates with no charge density on the Si dimers (i.e. pure bulk bands) are plotted with black dots. For comparison, conventional bulk bands, calculated at $k_z$ = 0 with a repeated cell consisting of four Si atoms, are superimposed as thick gray curves and labeled as $\Sigma_1$, $\Sigma_{1'}$ and  $\Sigma_2$. As might be expected, the 6 L slab calculation does not produce any bulk bands near the surface bands [Fig. \ref{fig:Band}(a)]. With increasing slab thickness, the bulk bands develop, and the VBM becomes dominated by a bulk state, instead of the DB state, when the slab thickness is greater than 30 L. Full agreement between bulk valence bands in slab and bulk calculations is only obtained for a slab thickness of 60 L or more. This thickness matches with that of the slab calculation in which the bulk band gap agrees with the value of a repeated cell bulk calculation, as seen in Fig. \ref{fig:PDOS} (b). 

In a slab calculation, where the electronic states in the sample have no periodicity perpendicular to the surface and Bloch's theorem does not apply, extra bands appear and develop with increasing slab thickness (in contrast to a bulk calculation which requires increasing BZ sampling for convergence).  The increase in number of bands and the development of their shape is seen clearly in Fig. \ref{fig:Band}. Some of the bands that appear with increasing thickness have charge density on both the Si dimers and the bulk atoms, as seen in Fig. \ref{fig:PARCHG} (b). These bands are hybridized between surface and bulk states. These hybridizations are prominent for the $\pi_1$ band, particularly near the $\Gamma$ point. As the slab thickness, and hence number of additional bands, becomes greater, more band mixing occurs: a thicker slab shows a larger number of hybridized bands. This causes the charge density of the $\pi_1$ band near the $\Gamma$ point to redistribute into the slab, resulting in an apparent absence of the $\pi_1$ band near the $\Gamma$ point, while the same band near the $\frac{1}{2}$J$^{\prime}\Gamma$ point, and the $\pi_2$ band across the BZ, are well localized at the surface regardless of slab thickness (See Fig.~\ref{fig:PARCHG_Z}). 

\begin{figure}
\includegraphics[width=12cm]{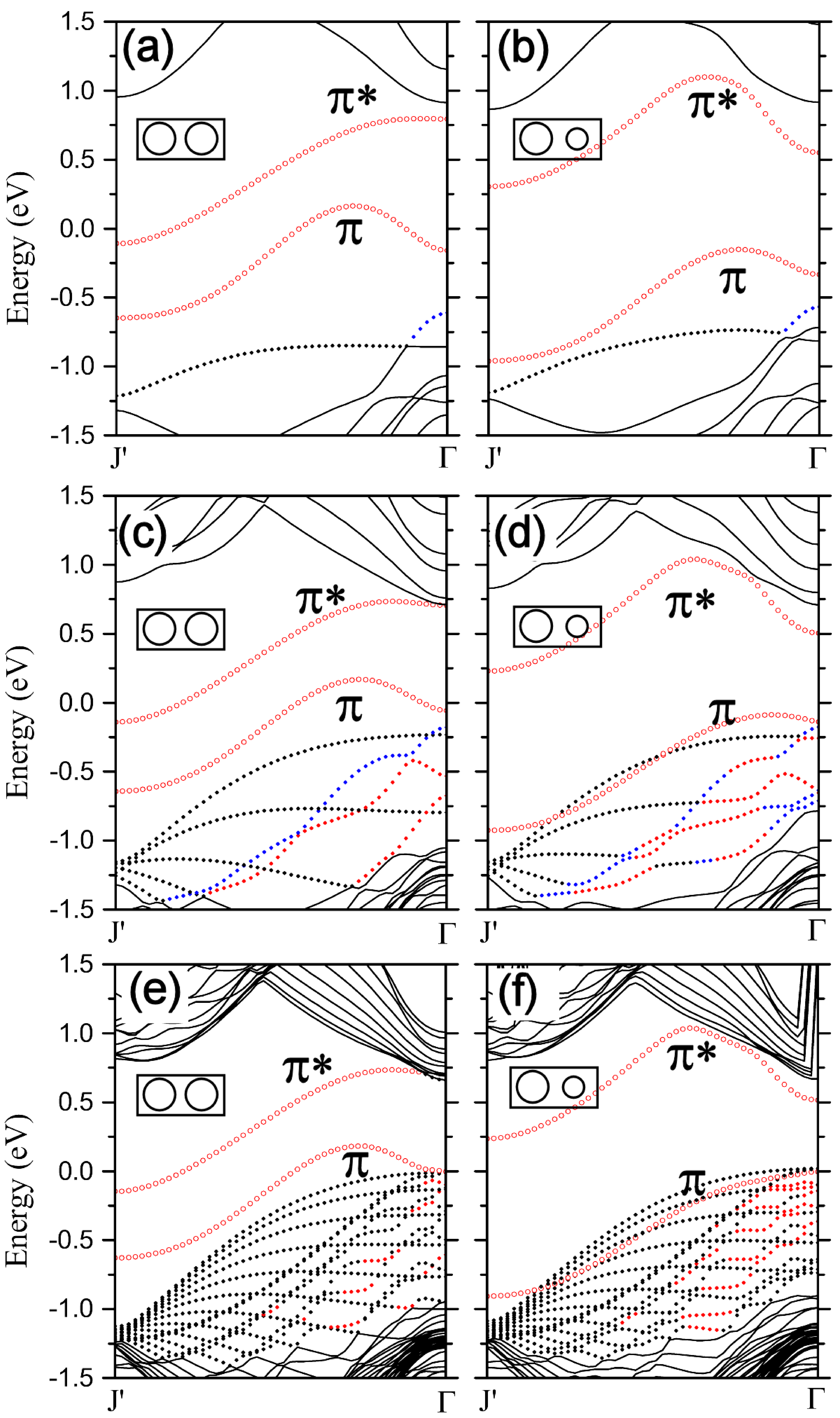} 
\centering
\caption{Band structures of (a), (c), (e) 2$\times$1 symmetric dimer and (b), (d), (f)  2$\times$1 asymmetric dimer surfaces obtained with slabs of: (a), (b) 6L;   (c), (d) 14L; (e), (f) 38L. Red open circle: DB states. red dots: hybrid states of  DB and bulk states, black dots: bulk states, solid line: unchecked bands.}
\label{fig:2x1Band}
\end{figure}

To investigate what causes this change in surface band energy and shape, we simplified the problem.  We calculated bands for 2$\times$1 symmetric (flat) and asymmetric (buckled) dimers with 6 L and 38 L slabs.  The only effect that causes the surface band to change shape is hybridization with the bulk bands. The $\pi$ band of the symmetric dimer in Fig. \ref{fig:2x1Band}(c) does not change its shape even though the slab thickness increases from 6 L to 38 L: since the $\pi$ band is situated in the bulk band gap, the hybridization is weak. On the other hand, the $\pi$ band of the asymmetric dimer changes its shape with slab thickness and this is not due to any structural change, but only due to the band mixing  between the $\pi$ and the bulk bands. The increase in slab thickness develops the bulk bands and raises their energy predominantly around the $\Gamma$ point; simultaneously, the hybridization lifts the $\pi$ band's energy near the $\Gamma$ point as well. 

These hybridizations are clearest between the $\Gamma$ point and a point halfway to the J$^{\prime}$ point. This tendency is kept when the bands are folded and transformed \emph{in the surface Brillouin zone} by switching the surface cell from 2$\times$1 periodicity to the \textit{p}(2$\times$2). Correspondingly, the $\pi_2$ band, which originates from the second half of the $\pi$ band in the smaller cell, does not change in its shape with slab thickness. Hybridization between surface and bulk bands is inevitable in slab calculations, so a precise calculation of surface states will require the convergence of bulk states. It is interesting to note that hybridized bands associated with bulk states, back bond (BB) and dimer bond (db) are also found in Fig. \ref{fig:Band}.   

\begin{figure}
\includegraphics[width=11cm]{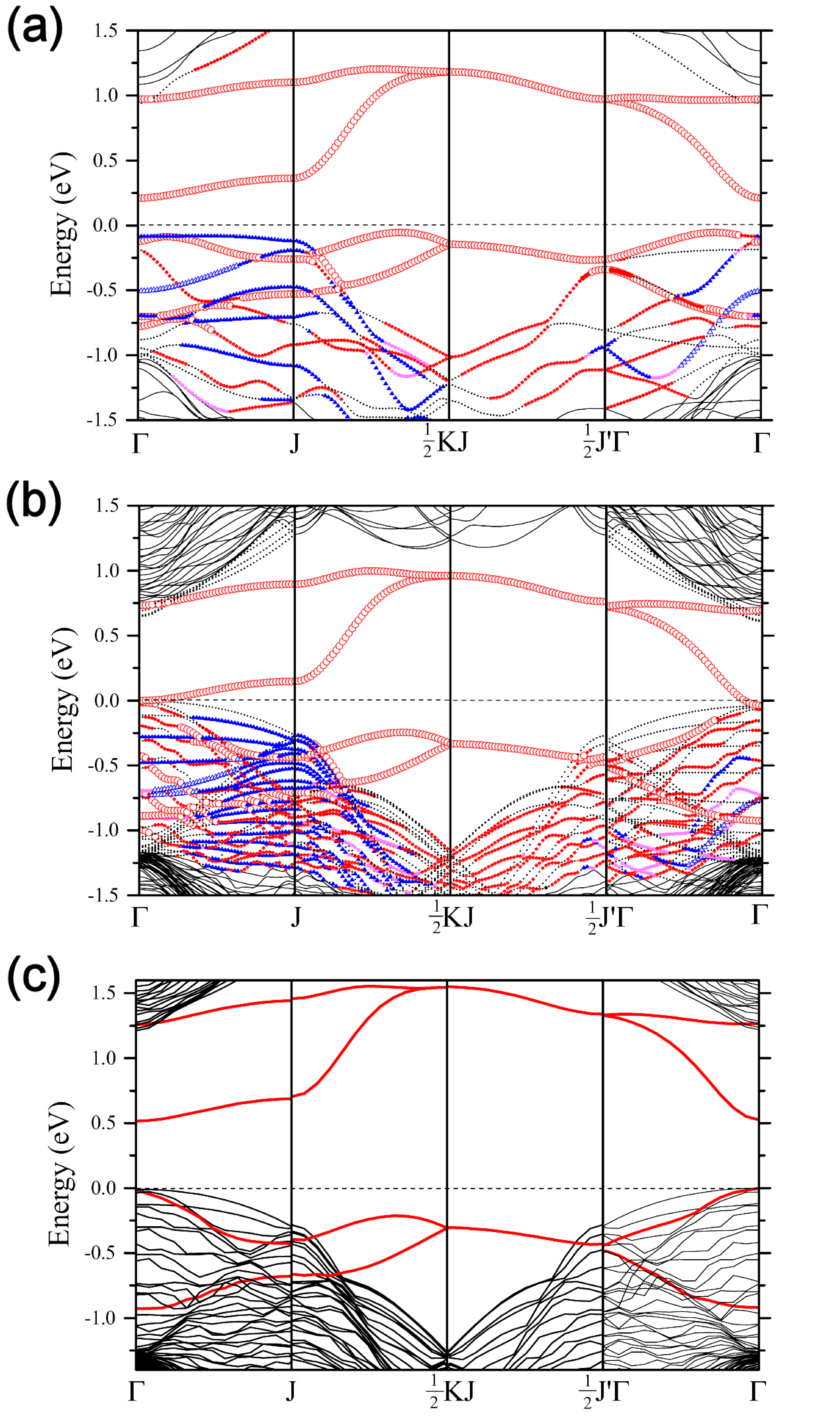}
\centering 
\caption{(Color online) Band structures of Si(100) \textit{p}(2$\times$2) calculated for slab thicknesses of (a) 10 L, (b) and (c) 38 L. (a) and (b) were calculated by GGA PBE functional and (c) by hybrid exchange-correlation functional (HSE06). To obtain the hybrid functional bands, both structural and electronic relaxations were also done by using HSE06. The usage of symbols are the same as that in FIG. \ref{fig:Band}. The DB states represented with red open circle are drawn in red lines in (c).}
\label{fig:10vs38}
\end{figure}

We have also examined the effect of slab thickness on the band structure along other high symmetry lines in the BZ. A comparison of band structures of Si(100) between 10 L and 38 L is displayed in Fig. \ref{fig:10vs38}. With a slab that is 10 L or thinner, as is typically used to model the Si(100) surface, the SBM tends to appear in the J-$\frac{1}{2}$KJ line [Fig. \ref{fig:10vs38}(a)]. This result can be found in previous reports \cite{Zhu89, Fritsch95, Ramstad95, Zhou13}. However, the SBM is actually located at the $\Gamma$ point in converged band structures calculated with a sufficiently thick slab [Fig. \ref{fig:10vs38}(b)]. This agrees with the observations of direct transition of the surface band gap \cite{Goldman86, Enta87}.

Due to the narrowing of the bulk band gap with the slab thickness, the bottom of the unoccupied DB band ($\pi_1^*$) overlaps the top of the valence band when the slab thickness is greater than 30 layers.  It might be that this overlap affects the development of the bands, so to check this, we carried out a band calculation for the 38 L slab with a screened hybrid exchange-correlation functional (HSE06) \cite{Heyd03, Heyd04, Heyd06}. These functionals generally alleviate the standard DFT problem with band gaps that are too small.  The results are shown in Fig. \ref{fig:10vs38}(c), where it is clear that the band gap has opened. There are no other differences in the band structure between the conventional and hybrid functional calculations.  Our discussions, based on standard DFT, are therefore valid and unaffected by the band gap. 

\section{Conclusion}
We have shown that DFT calculations for the electronic structure of the Si(100) surface require a much thicker slab than is conventionally used. A study for simple geometrical structures will be adequate with a thin slab of around 15 layers, though 20-25 layers are required for convergence of the surface energy. Treatment of the precise surface electronic structure needs a slab of at least 30 layers, while the full convergence of both surface and bulk electronic structure requires a slab of 60 layers or more. It is important to determine which properties need to be calculated and choose an appropriate slab size. With advances in experimental techniques, the desired accuracy from DFT modeling has reached an energy resolution in the meV range, suggesting that calculations should be performed with a significantly thicker slab than is the norm.

\ack{This work was in part supported by MEXT KAKENHI (24510160).}

\appendix

\section{Extra Figures and Tables}

The table and figures below are not central to our study, but add extra detail in some areas, and are included for completeness and information.

\begin{table}[h]
\caption{\label{tab:table2} Band width of $\pi_1$ and $\pi_2$ bands in the $\Gamma$-$\frac{1}{2}$J$^{\prime}\Gamma$ line.} 
\begin{tabular}{ccc}
	Slab thickness (L) & Bandwidth (eV) \\
    & $\pi_1$ (eV) & $\pi_2$ (eV) \\
	\hline
	6 & 0.130 & 0.434\\
	14 & 0.312 & 0.411\\
	22 & 0.385 & 0.434\\
	30 & 0.399 & 0.426\\
	38 & 0.399 & 0.416\\
	62 & 0.400 & 0.405\\
	78 & 0.410 & 0.405\\
\end{tabular}
\end{table}

\begin{figure}
\includegraphics[width=8.5cm]{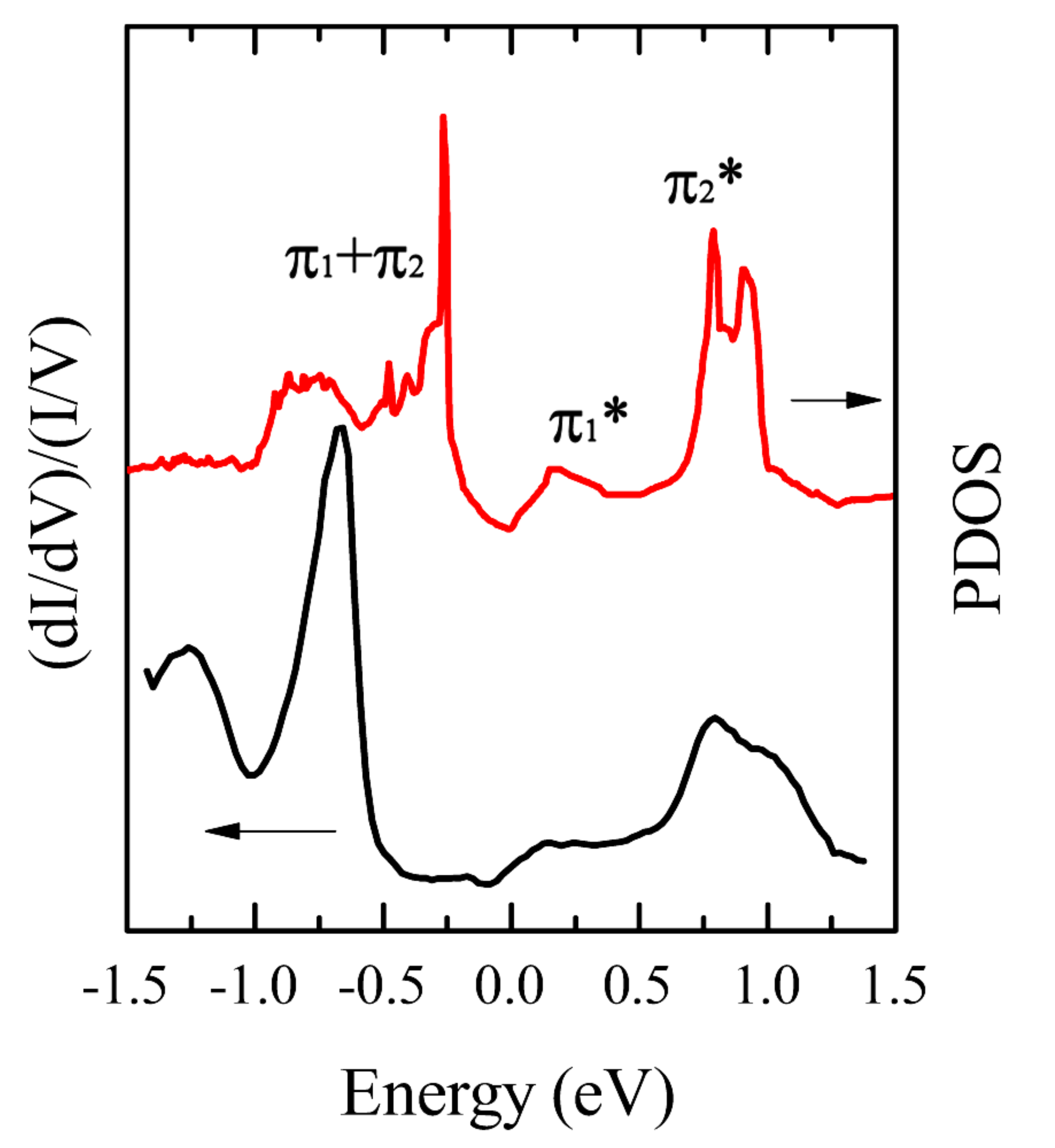} 
\caption{Tunneling spectrum and DFT-PDOS of Si dimer on the Si(100) surface. Tunneling spectrum was obtained at 78 K with a setpoint of Vs=+1.5 and I=3.5nA. PDOS was obtained with a 62 L slab.}
\label{fig:STSvsDFT}
\end{figure}

\begin{figure}
\includegraphics[width=13cm]{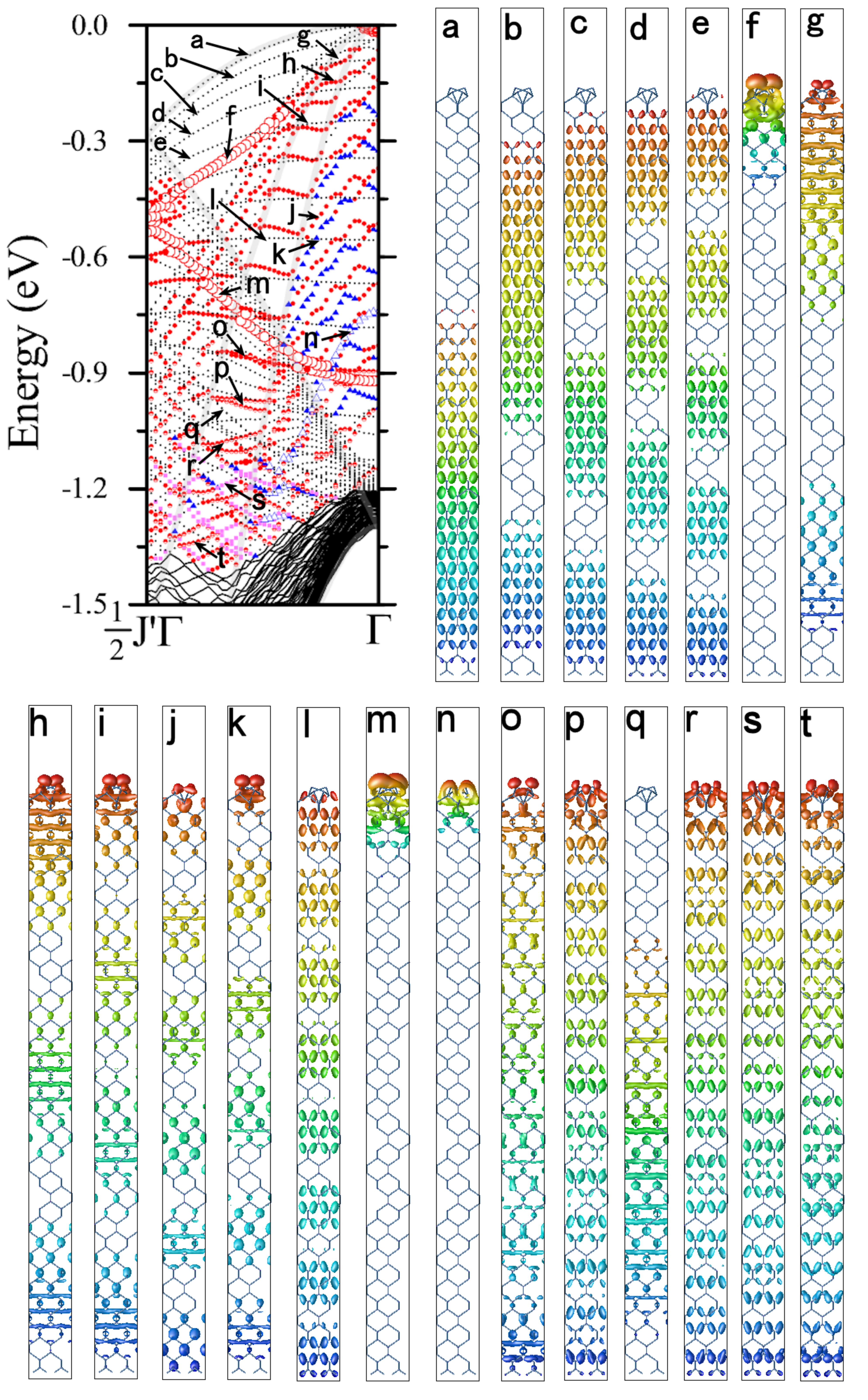} 
\caption{Isosurface charge density distributions of eigenstates indicated by labels in the band diagram of Si(100)-\textit{p}(2 $\times$ 2)-78 L. Isosurface value is 1$\times$10$^{-3}$ e/\AA $^3$.}
\label{fig:PARCHG78}
\end{figure}

\begin{figure}
\includegraphics[width=15cm]{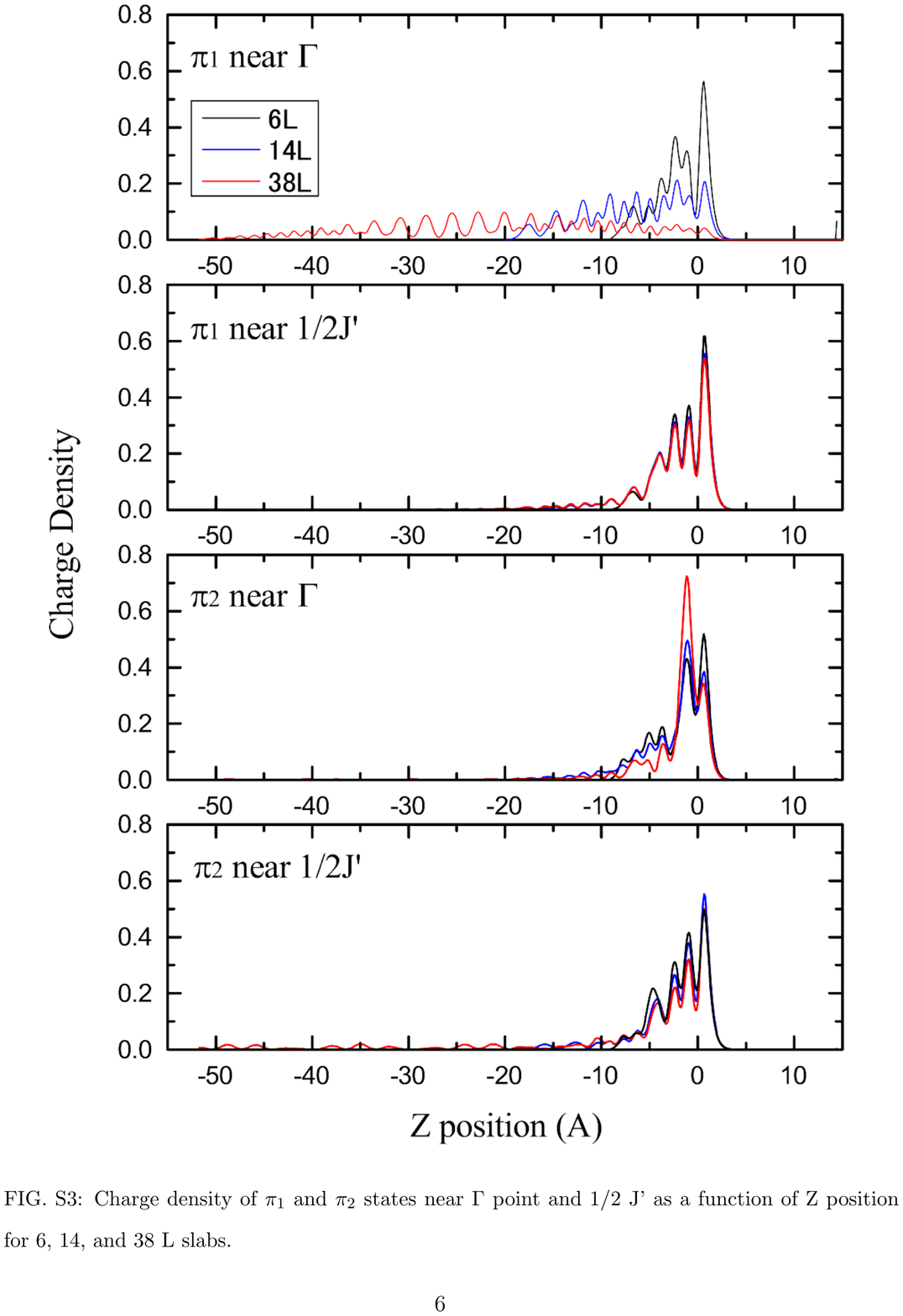} 
\caption{Charge density of $\pi_1$ and $\pi_2$ states near $\Gamma$ point and 1/2 J' as a function of Z position for 6, 14, and 38 L slabs.}
\label{fig:PARCHG_Z}
\end{figure}

\end{document}